\documentclass[aps,prl,twocolumn,groupedaddress]{revtex4}
\usepackage{graphicx}
\begin{document}


\title{Effect of Ga$^{+}$ irradiation on magnetic and magnetotransport properties \\in (Ga,Mn)As epilayers }

\author{H. Kato, K. Hamaya$^{a)}$, Y. Kitamoto, T. Taniyama,$^{1}$ and H. Munekata$^{2}$}

\affiliation{Department of Innovative and Engineered Materials, Tokyo Institute of
Technology,\\ 4259 Nagatsuta, Midori-ku, Yokohama 226-8502, Japan.\\
$^{1}$Materials and Structures Laboratory, Tokyo Institute of
Technology,\\ 4259 Nagatsuta, Midori-ku, Yokohama 226-8503, Japan.\\
$^{2}$Imaging Science and Engineering Laboratory, Tokyo Institute of Technology,\\ 
4259 Nagatsuta, Midori-ku, Yokohama 226-8503, Japan.}




\date{\today}
\begin{abstract}
We report on the magnetic and magnetotransport properties of ferromagnetic semiconductor (Ga,Mn)As modified by Ga$^{+}$ ion irradiation using focused ion beam. A marked reduction in the conductivity and the Curie temperature is induced after the irradiation. Furthermore, an enhanced negative magnetoresistance (MR) and a change in the magnetization reversal process are also demonstrated at 4 K. Raman scattering spectra indicate a decrease in the concentration of hole carriers after the irradiation, and a possible origin of the change in the magnetic properties is discussed. 

${}^{a)}$  e-mail: hamaya@iem.titech.ac.jp
\end{abstract}


\maketitle

Hole-mediated ferromagnetic properties in III-V magnetic semiconductors have been intensively studied from theoretical and experimental points of view.\cite{Ohno2,Dietl2,Oiwa,Matsukura,Keavney,Abolfath} In order to use these materials in spin electronic devices, a promising approach to manipulating the hole-mediated properties is now required. In this study, we focus on Ga$^{+}$ ion irradiation technique which has been used for electronic modification of GaAs.\cite{Hirayama,Nakata} Since Ga$^{+}$ ion irradiation generates deep trap levels,\cite{Dansas} Ga$^{+}$ ion irradiation technique using focused ion beam (FIB) is expected to locally change the hole concentration of ferromagnetic $p$-(Ga,Mn)As and the magnetic properties.

In this article, we report on a clear experimental demonstration of modifying the ferromagnetic properties of (Ga,Mn)As using Ga$^{+}$ ion irradiation. The Curie temperature and conductivity are reduced after the irradiation. Also, the magnetoresistance (MR) under high magnetic fields is enhanced significantly, indicating an increase in the magnetic fluctuation of (Ga,Mn)As. Analyses of Raman scattering spectra reveal that the Ga$^{+}$ ion irradiation induces a marked reduction in the hole carriers, which is attributed to the formation of deep trap levels. 

A 100-nm-thick Ga$_{0.946}$Mn$_{0.054}$As epilayer was grown on a GaAs(001) substrate using low-temperature molecular beam epitaxy at 235$^{\circ}$C.  For transport measurements, the epilayer was patterned into a rectangular bar structure (150 $\mu$m $\times$ 10 $\mu$m) with a long axis along [100] with Ti/Au ohmic contacts using a combination of electron beam lithography, wet-etching and lift-off method. After magnetotransport measurements of a non-irradiated sample, an area between the two voltage terminals of the sample (10 $\mu$m $\times$ 10 $\mu$m) was irradiated by Ga$^{+}$ ion beam using HITACHI FB-2000A; a schematic diagram of the irradiated sample is illustrated in Fig. 1. Ion irradiation was carried out at an acceleration voltage of 30 kV and a beam dose density of 4.0 $\times$ 10$^{15}$ ion/cm$^{2}$. To investigate the crystallinity and surfaces morphology in the irradiated area, we used a high-resolution x-ray diffractomeater (HR-XRD) and an atomic force microscope. Magnetotransport measurements were performed using a standard four-point ac method and a magnetic field was applied parallel to the current flow ([100]). To gain an insight into the effect of ion irradiation, we measured Raman scattering spectra using Ar$^{+}$ laser at a wevelength of 514.5 nm.
\begin{figure}[b]
\includegraphics[width=6cm]{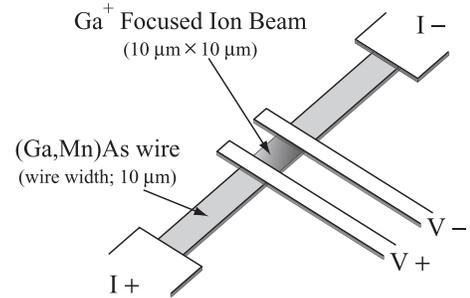}
\caption{Schematic diagram of the patterned sample with irradiation of Ga$^{+}$ ion beam.}
\end{figure}

Figure 2 (a) shows the temperature-dependent resistance ($R-T$ curve) of both non-irradiated sample and irradiated sample in $H =$ 0 Oe and 90 kOe. Typical $R-T$ curves are observed for the non-irradiated sample\cite{Matsukura} ; the $R-T$ curve in $H =$ 0 Oe shows a maximum ($T_\mathrm{p}$) at 70 K which is close to the Curie temperature ($T_{c}$) of this sample. A negative magnetoresistance (MR) is seen in $H =$ 90 kOe, and the largest MR occurs due to a large magnetic fluctuation at $T_\mathrm{p}$. For the irradiated sample, the conductivity is markedly reduced and the $R-T$ curve changes from a metallic behavior to an insulating behavior. Also, a negative MR in $H =$ 90 kOe is significantly enhanced at a low temperature regime ($T$ $<$ 50 K), compared with that of the non-irradiated sample. Defining MR ({\%}) as $[(R_{H}- R_{0}) / R_{0}]$ $\times$ 100, where $R_{H}$ and $R_{0}$ are the resistance in a magnetic field of {\it H} and zero field, respectively, the negative MR ({\%}) in $H =$ 90 kOe increases from $-$ 13.7 {\%} to $-$ 67.6 {\%} at 4 K after irradiation.   
\begin{figure}[t]
\includegraphics[width=8cm]{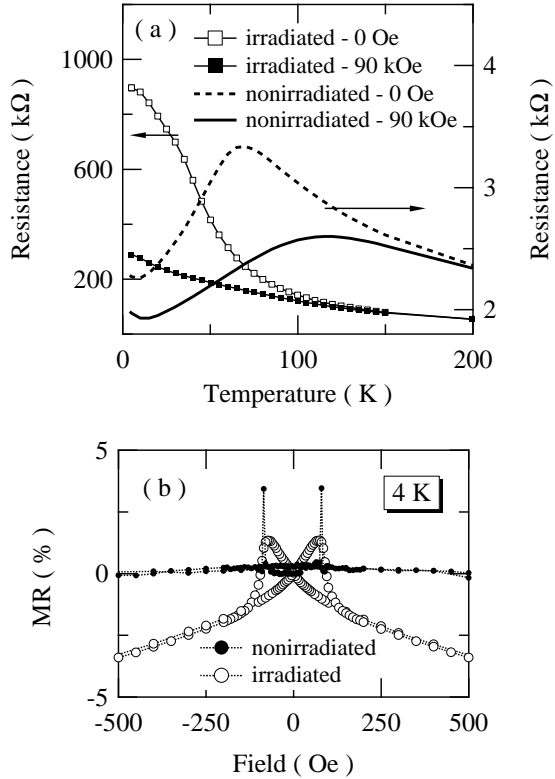}
\caption{( a ) The temperature-dependent resistance and ( b ) field-dependent magnetoresistace of irradiated sample and non-irradiated sample in zero field and a magnetic field of 90 kOe. }
\end{figure}

Figure 2 (b) shows the field-dependent MR of both samples in a low-field regime at 4 K. The hysteresis curve of the non-irradiated sample is similar to those reported in our previous works,\cite{Hamaya2,Kato} where the magnetization reversal process is governed by 90$^{\circ}$ domain wall displacement associated with $\left\langle 100 \right\rangle$ cubic magnetocrystalline anisotropy. Abrupt MR jumps at $H =$ $\pm$ 85 Oe correspond to magnetization switching towards a direction perpendicular to the current flow, i.e., [010]. These features clearly indicate a rapid magnetization reversal via domain wall displacement. In contrast, the hysteresis curve of the irradiated sample is changed dramatically. In particular, the magnetization reversal process is likely to be governed by magnetization rotation: no abrupt magnetization switching occurs although two MR peaks are seen at the same field of $\pm$ 85 Oe. With increasing temperature, the hysteresis behavior disappears at 35 K (not shown), implying that $T_{c}$ of the irradiated sample is approximately 35 K. 

To examine the origin of the effect of Ga$^{+}$ ion irradiation, we measure micro-Raman spectra of the irradiated sample (solid curve) and the non-irradiated sample (dashed curve) at room temperature (Fig. 3). Prior to the measurement of Raman spectra, we confirmed that $\sim$ 2 nm from the surface was etched by the ion irradiation. Since HR-XRD revealed that the crystal structure became disordered by the irradiation down to 30 nm from the surface and the remaining 70-nm-region has the same crystal structure as the non-irradiated sample, the Raman spectra were collected for both samples after etching 50 nm from the surface to eliminate an influence from the disordered region. The spectrum of the non-irradiated sample shows two peaks even after the etching as reported in a previous study of (Ga,Mn)As.\cite{KCKu} Since probing light penetrates into the GaAs buffer layer, the peak at 295 cm$^{-1}$ originating from GaAs bulk is significant. For the spectrum of the irradiated sample, on the other hand, a significant decrease in the peak intensity near 270 cm$^{-1}$ is clearly seen and a small increase in the peak near 295 cm$^{-1}$ is also observed. 
\begin{figure}[t]
\includegraphics[width=7cm]{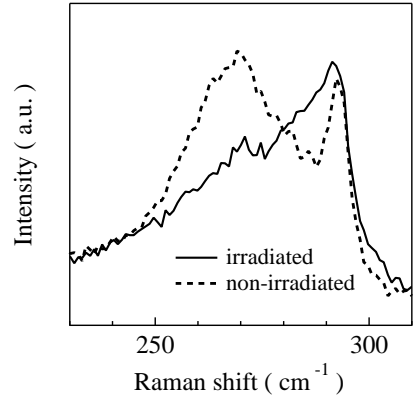}
\caption{Raman spectra of a (Ga,Mn)As epilayer before (dashed curve) and after (solid curve) irradiation. The measurements were done after etching 50-nm-surface region of the epilayer.}
\end{figure}

To assign the two peaks at 270 cm$^{-1}$ and 295 cm$^{-1}$, we examine the Raman spectra of a semi-insulating GaAs ($i$-GaAs) wafer and a $p$-GaAs (Zn-dope: $p \sim$ 10$^{19}$ cm$^{-3}$) wafer before and after irradiation as shown in Fig. 4. The surface of the irradiated samples were also etched by 50 nm before Raman measurements. For non-irradiated $i$-GaAs, two peaks are observed at 265 cm$^{-1}$ and 290 cm$^{-1}$ as seen in the (Ga,Mn)As samples. These peaks are assigned to transverse optical (TO) phonon mode and longitudinal optical (LO) phonon mode at $\Gamma$ point as in literatures, respectively.\cite{Tiong} After irradiation, the basic feature of both TO-phonon and LO-phonon modes does not change, although the peak intensities are slightly decreased. This result indicates that a region with a destroyed crystal structure due to irradiation is removed by pre-etching the surface.\cite{Ishioka} On the other hand, the feature of non-irradiated $p$-GaAs is significantly different from those of $i$-GaAs: the peak at 265 cm$^{-1}$ is larger than that at 290 cm$^{-1}$. Since the hole concentration is 10$^{19}$ cm$^{-3}$, plasma oscillation likely occurs in $p$-GaAs and a coupled plasmon$-$LO-phonon mode (CPLOM) appears being overlapped with TO-phonon mode at 265 cm$^{-1}$,\cite{Irmer} and the CPLOM reduces free LO-phonons at 290 cm$^{-1}$, accordingly.\cite{Yuasa} After irradiation, CPLOM is reduced so that the LO phonon mode is recovered, giving rise to a spectrum similar to that of  $i$-GaAs. This suggests that Ga$^{+}$ ion irradiation reduces the hole carrier and the corresponding plasma oscillation in $p$-GaAs. 
\begin{figure}[t]
\includegraphics[width=6cm]{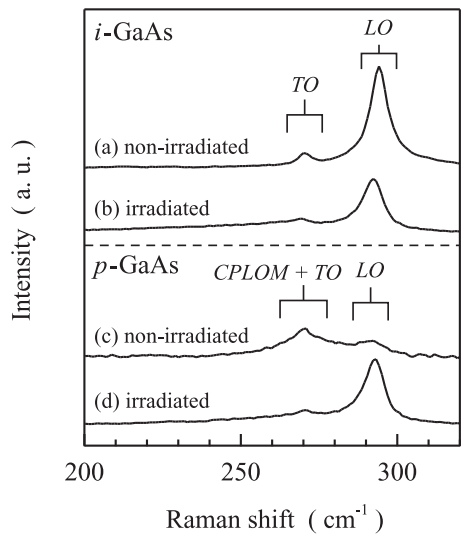}
\caption{Raman spectra of $i$-GaAs and $p$-GaAs wafers  before and after ion irradiation. The measurements were done after etching 50-nm-surface region of the samples.}
\end{figure}

From the results of Ga$^{+}$ ion irradiation for $p$-GaAs, we consider a possible origin of the irradiation effect on the Raman spectrum of (Ga,Mn)As in Fig. 3. We can clearly notice that the fact that ion irradiation on (Ga,Mn)As reduces the CPLOM at 265 cm$^{-1}$ and increases LO phonon mode slightly is similar to the results of $p$-GaAs in Fig. 4. This indicates that the hole carriers in (Ga,Mn)As decrease after irradiation. A cause of the reduction of hole carriers may be carrier trapping at trap levels in the band gap of (Ga,Mn)As generated by Ga$^{+}$ ion irradiation. This description clearly gives a comprehensive understanding of the magnetic and magnetotransport properties shown in Fig. 2. Since the ferromagnetism in (Ga,Mn)As occurs via $p$-$d$ exchange interaction, the reduction in the hole carrier should cause instability of the ferromagnetism, leading to a reduction in the $T_c$ and thermal fluctuation even at low temperature. The thermal fluctuation gives an enhancement in the negative MR in 90 kOe and changes MR curves as observed in this study. The magnetization data are thus consistent with our description that hole carriers are trapped and the hole concentration is much reduced by ion irradiation. These result  indicate that modification of the ferromagnetism of (Ga,Mn)As can be achieved utilizing ion irradiation.

We have demonstrated Ga$^{+}$ ion irradiation on a (Ga,Mn)As epilayer using focused ion beam. The ion irradiation brought  about an enhancement in the negative MR and a change in the magnetization reversal process. Micro-Raman measurements explain a decrease in the free hole carriers by ion irradiation. The reduction of hole carriers affects $p$-$d$ exchange interaction in (Ga,Mn)As, resulting in modification of the magnetic properties.

\begin{acknowledgments}
This work was supported in part by the Scientific Research in Priority Areas ``Semiconductor Nanospintronics" of the Ministry of Ecucation, Culture, Sports, Science and Technology, Japan. 
The authors gratefully acknowledge Dr. K. Saito of BRUKER AXS K. K. for measurements of micro-X-ray diffraction, and are grateful to Prof. Y. Yamazaki of the Tokyo Institute of Technology (T. I. Tech.) for kindly offering the opportunity to use FIB. The stimulating discussion comments on Raman spectra by Prof. K. G. Nakamura of T. I. Tech. are greatly appreciated.
\end{acknowledgments}


\end{document}